\documentclass{PoS}
\usepackage{url}

\title{The Integrable Optics Test Accelerator}

\ShortTitle{The Integrable Optics Test Accelerator}

\author{\speaker{B. Freemire} \thanks{Work supported by the U.S. Department of Energy, Office of High Energy Physics, under Contract Nos. DE-AC02-07CH11359 and DE-AC02-05CH1123 and General Accelerator Research and Development (GARD) Program.} \\
        Northern Illinois University, DeKalb, Illinois 60115, USA \\
        E-mail: \email{bfreemire@niu.edu}}

\author{J. Eldred \\
        Fermi National Accelerator Laboratory, Batavia, Illinois 60510, USA \\
        E-mail: \email{jseldred@fnal.gov}}

\author{for the FAST/IOTA Team}

\abstract{
The Integrable Optics Test Accelerator (IOTA) at the Fermilab Accelerator Science \& Technology (FAST) Facility is beginning operations, with an experimental program aimed at developing technology to enable future high intensity particle beams.
The results garnered from IOTA should allow for better control of beam losses, leading to higher safely achievable beam power.
This will be particularly useful for future accelerator based neutrino experiments.
Two of the R\&D efforts at IOTA, Nonlinear Integrable Optics and Electron Column-based space charge compensation, are discussed here.
}

\FullConference{The 20th International Workshop on Neutrinos (NuFact2018)\\
		12-18 August 2018\\
		Blacksburg, Virginia}

\begin{document}

\section{Introduction}

\begin{figure}
    \centering
    \includegraphics[height=95pt]{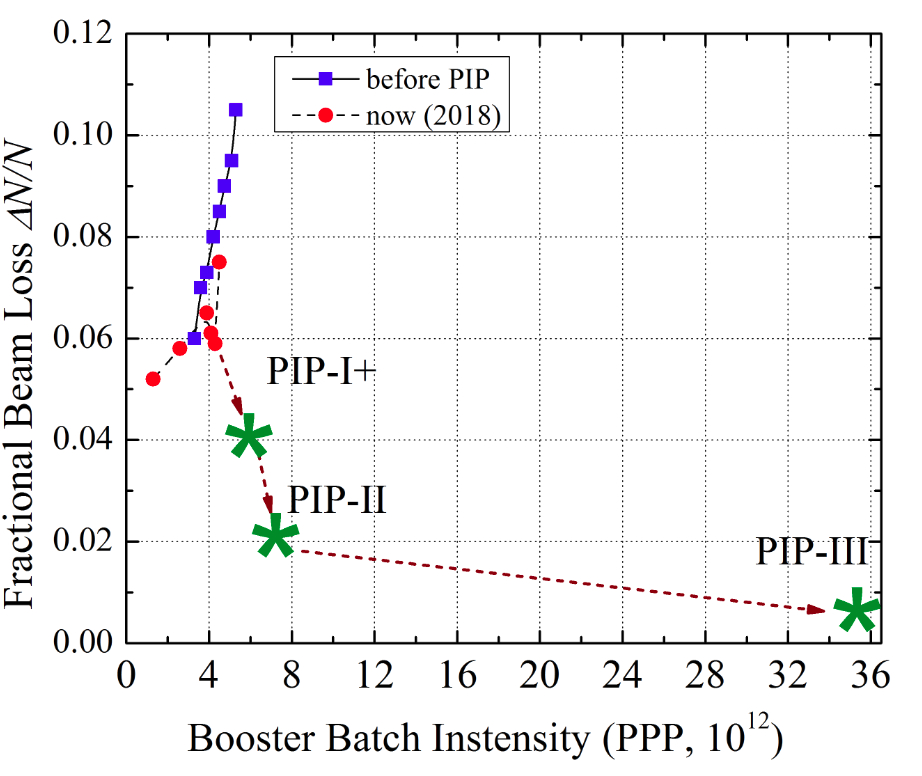} \quad \quad \quad
    \includegraphics[height=95pt]{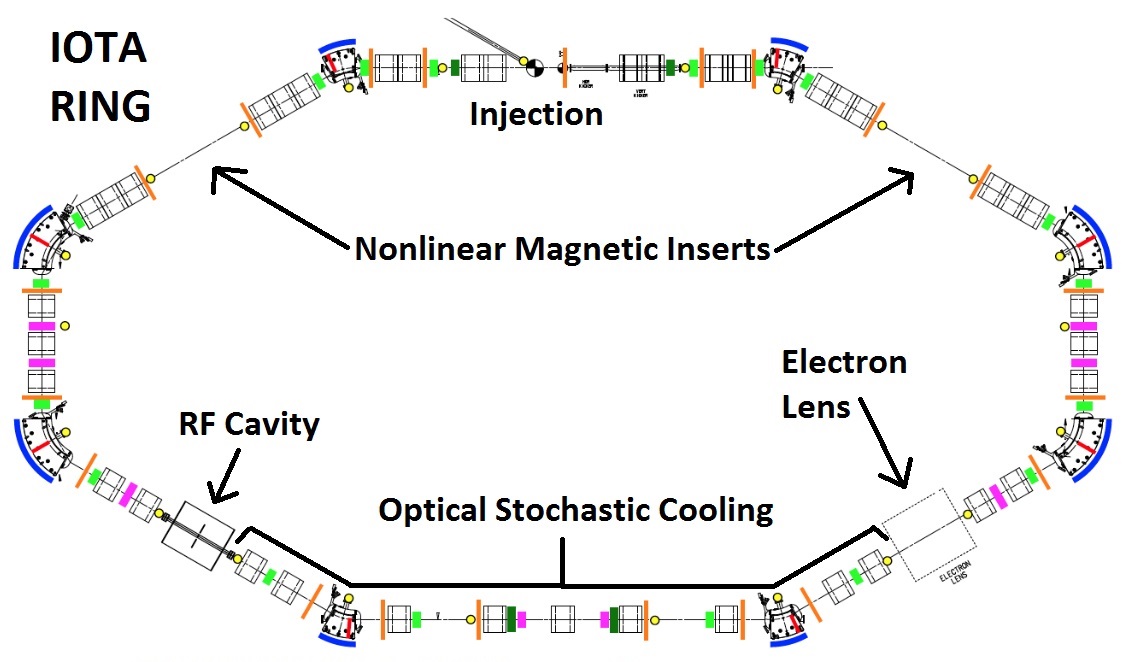}
    \caption{Left - Fractional beam loss vs. protons per pulse intensity in the Fermilab Booster~\cite{Shiltsev}. Right - Diagram of the IOTA ring, showing the locations of the Nonlinear Magnet(s) and Electron Lens/Column~\cite{AntipovIOTA}}
    \label{fig:1}
\end{figure}

The next generation of accelerator-based neutrino experiments rely on proton drivers exceeding the 1 MW benchmark~\cite{LBNF,T2HK,ESSnuSB}.
At Fermilab, the Proton Improvement Plan II (PIP-II) will provide 1.2 MW beam power at 120 GeV, which is the initial requirement for the Deep Underground Neutrino Experiment (DUNE) physics program~\cite{DUNE}. 
The DUNE baseline calls for a further upgrade to 2.4 MW~\cite{DUNEidr} while simultaneously decreasing the fractional beam loss in the first ring of the accelerator complex, the Booster (see the left plot in Fig.~\ref{fig:1}). 
A new rapid-cycling synchrotron (RCS) has been proposed to replace the Booster~\cite{Shiltsev,EldredPAC,Prebys}.

An accelerator R\&D program has been launched at the Fermilab Accelerator Science \& Technology (FAST) Facility with the goal of demonstrating technology for the next generation of particle accelerators. 
The current flagship of FAST's experimental program is the Integrable Optics Test Accelerator (IOTA), a modular electron/proton storage ring~\cite{AntipovIOTA}.
Two technologies in particular, Nonlinear Integrable Optics and Electron Lens/Column space-charge compensation, are being developed at IOTA to enhance the performance of high-intensity hadron rings~\cite{Eldred18,Freemire18}. The layout of IOTA is shown in Fig.~\ref{fig:1} on the right.

\section{Nonlinear Integrable Optics}

Nonlinear Integrable Optics is a development in particle accelerator technology that enables strong nonlinear focusing without generating new parametric resonances~\cite{Danilov}. Integrable optics is considered for high-intensity rings, where nonlinearity is known to suppress halo formation~\cite{WebbArxiv} and enhance Landau damping of charge-dominated collective instabilities~\cite{Stern}.

\begin{figure}
    \centering
    \includegraphics[height=90pt]{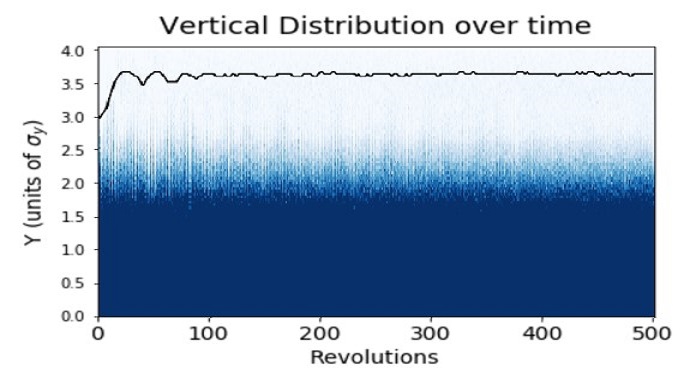}
    \includegraphics[height=90pt]{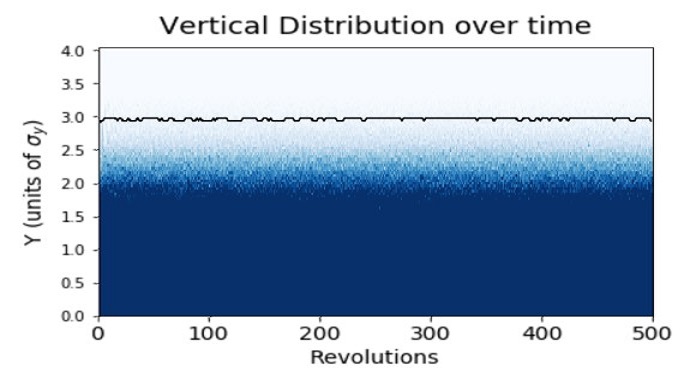}
    \caption{Synergia space-charge simulation of an intense mismatched waterbag beam in an integrable RCS lattice~\cite{EldredCBB}. Left - Linear Lattice. Right - Nonlinear Integrable Lattice.}
    \label{fig:2}
\end{figure}

In high-intensity beams, small disturbances of the beam core can cause outlying particles to be drawn into the beam halo~\cite{Wangler}. Under conditions of strong nonlinear focusing, the mismatched particle beams rapidly reach an equilibrium distribution that can suppress halo~\cite{Hall}. Simulations of nonlinear integrable lattices demonstrate that halo can be suppressed (see Fig.~\ref{fig:2}) while the large betatron tune-spread is not impacted by crossing fourth-order resonance lines~\cite{WebbArxiv,Eldred17}.

Perfect integrable motion requires particles to have precisely defined phase-advances, but the phase-advances would be shifted by the space-charge forces. The lattice is tuned to compensate the linear space-charge tune-shift at a certain beam intensity~\cite{Romanov} and a recent simulation of an integrable RCS lattice showed stable nonlinear motion with a space-charge tune-shift of -0.4~\cite{Eldred18}.


Webb et al.~\cite{Webb1,Webb2} showed that chromaticity of an integrable lattice does not have to be corrected to zero; the motion of an off-momentum particle is integrable if the horizontal and vertical chromaticity are matched. For a ring composed of several integrable cells the synchrotron motion is adiabatic~\cite{Webb2,Webb3} and the beam can be stable under significant chromatic tune spread~\cite{Eldred18}.

\section{Space Charge Compensation}

\begin{figure}
    \centering
    \includegraphics[height=95pt]{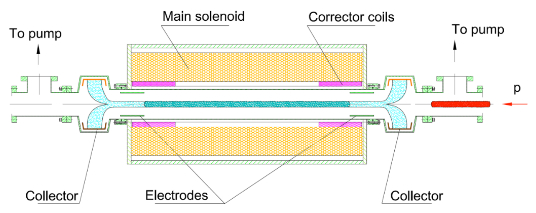} \quad \quad
    \includegraphics[height=95pt]{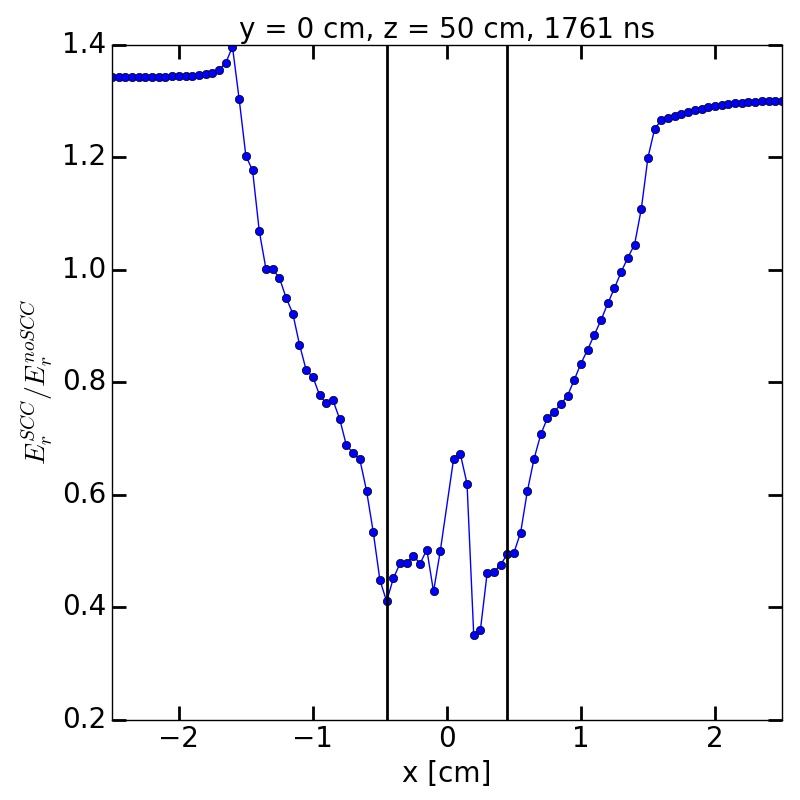}
    \caption{Left - Cutaway view of the Electron Column.  Right - Ratio of the radial electric field with the Electron Column to that of the beam self-field, indicating space charge compensation within the radius of the beam (marked)~\cite{FreemireHB}.}
    \label{fig:3}
\end{figure}

An Electron Column will be tested in IOTA that is similar to an Electron Lens, but simpler in that no electron gun or collector, or transport magnets are required.
The Electron Column works by matching the profile of a plasma created by ionization of a short section of (hydrogen) gas to that of the beam using electrodes and a solenoid (see the diagram on the left in Fig.~\ref{fig:3}).
The solenoid must be strong enough to suppress electron-proton instabilities, but weak enough to allow the ions produced to escape the Column, leaving the electrons to neutralize the charge of the beam.

Simulations of the Electron Column show space charge compensation within the beam's radius after a single turn, as shown in Fig.~\ref{fig:3} on the right~\cite{Freemire18}.
Current studies are focused on following the evolution of the plasma and a beam matched to the rest of the IOTA lattice over many turns.

\section{Outlook}

Commissioning of the Nonlinear Magnet in IOTA with electrons has recently begun. 
After the first round of electron experiments, a proton source and RFQ will be installed, and commissioning with protons in IOTA will take place. 
IOTA will focus on innovative results for the production of more intense beams, at Fermilab and at large.


\end{document}